\documentclass[12pt]{article}
\usepackage{graphicx}
\newcommand{\sss}{\scriptscriptstyle}

\newcommand {\be}{\begin{equation}} 
\newcommand{\ee}{\end{equation}}    
\def\ddt{\frac{\partial}{\partial t}}

\def\dds1{\frac{\partial}{\partial s_1}}

\def\vte{v_{{\sss T}e}}
\def\vtj{v_{{\sss T}j}}
\def\vta{v_{{\sss T}a}}
\def\vtb{v_{{\sss T}b}}

\def\d{d\kern-0.8 ex\vrule height 1.3 ex depth-1.24 ex width 0.7 ex
\kern 0.15 ex}
\def\D{D\kern-1.7 ex\vrule height .87 ex depth-0.8 ex width 0.7 ex
\kern 0.95 ex}

\begin{document}
\baselineskip 20 pt

\begin{center}

\Large{\bf On electrostatic modes in multi-ion and pair-ion  collisional plasmas }

\end{center}

\vspace{0.7cm}

\begin{center}

 J. Vranjes

{\em Center for Plasma Astrophysics, Celestijnenlaan 200B, 3001 Leuven,
 Belgium,
and
Facult\'{e} des Sciences Appliqu\'{e}es, avenue F.D. Roosevelt 50,
 1050 Bruxelles, Belgium.}

\vspace{5mm}

D. Petrovic,

{\em Research Group PLASMANT, Department of Chemistry, University of Antwerp, Universiteitsplein 1,
2610 Antwerp, Belgium
}

\vspace{5mm}

B. P. Pandey

  {\em  Department of Physics, Macquarie University, Sydney, NSW
2109, Australia.}

\vspace{5mm}

S. Poedts

{\em Center for Plasma Astrophysics, and Leuven Mathematical Modeling and Computational Science Centre
 (LMCC) Celestijnenlaan 200B, 3001 Leuven,  Belgium.
}

\end{center}

\vspace{2cm}

{\bf Abstract:} The physics of plasmas containing positive and negative ions is discussed with special attention to the recently produced pair-ion plasma containing ions of equal mass and opposite charge. The effects of the density gradient in the direction perpendicular to the ambient magnetic field vector, observed in the experiment, are discussed. In addition, the possible
presence of electrons is discussed in the context of plasma modes propagating  at an angle with respect to the magnetic field vector. The electrons may either be added to the plasma or enter the plasma attached to negative ions and then become released from the ions as a result of collisions.  It is shown that the electron plasma mode in a cold plasma may become a backward mode in the presence of a density gradient, and this behavior may be controlled either by the electron number density or the mode number in the perpendicular direction. In plasmas with hot electrons an instability may develop, driven by the combination of electron collisions and the density gradient, and in the regime when the ions' response is similar to a sound mode, i.e., for  un-magnetized ions, implying  mode frequencies much above the ion gyro-frequency or mode wave-lengths shorter than the ion gyro-radius. In the case of a pure pair-ion plasma, for lower frequencies  and for parameters close to those used in the recent experiments, the perturbed ions may feel the effects of the magnetic field. In this case the plasma mode also becomes backward, resembling features of an  experimentally observed but yet unexplained backward mode.

\vspace{2cm}


\vspace{2cm}

\pagebreak

\vspace{0.7cm}

\noindent{\bf I. \,\,\, INTRODUCTION}

\vspace{0.7cm}

Space plasmas only exceptionally include a single ion species. Much more frequently, they consist of a mixture of several ion species with different charges and degrees of ionization, different masses and concentrations, on a background of usually more mobile electrons. In the laboratory, the situation is not much different. A particularly intriguing laboratory example of a multi-ion plasma  is the case of the so-called pair-ion plasma. The term 'pair-ion' implies a plasma containing oppositely charged ions with nearly the same mass. In the recent past, there has been a lot of activity on the theory and the experiments dealing with such plasmas, initiated by a series of works describing the experimental production of pair ions.$^{1-5}$ The ions in question are in fact  $C_{60}^{\pm}$, produced in simultaneous processes of impact ionization
and electron attachment, that are further collected by a magnetic filtering effect, i.e., by a  diffusion in the radial direction (perpendicular to the magnetic field lines). In this way the ions are first separated from the electrons, and then collected through an annular hole (with 3~cm diameter),  due to a subsequent downstream along the magnetic lines, in a very narrow and elongated chamber (90~cm in length, and with an 8~cm diameter).
In Refs.~2-5 several types of modes have been reported, e.g., the ion plasma wave (IPW), the ion acoustic wave (IAW), and the so-called intermediate frequency (IF) wave. Detailed measurements presented in the most recent Refs.~4 and 5, reveal that the IAW  actually has two separate branches, accompanied by some additional backward propagating mode situated between them.

The works 1-5 have attracted a lot of interest from researchers. This has resulted in numerous theoretical studies$^{6-15}$ dealing with various aspects of linear and nonlinear waves and instabilities in pair plasmas. The IF wave has been particularly in focus in a number of works, because the behavior of this mode does not fit into the standard plasma theory. Yet, the IAW also seems to be problematic because it should not appear in the perfectly anti-symmetric pair-ion plasma in question here (i.e., with two ion groups with equal masses and temperatures, but opposite charges).
However, specific conditions in the experiments$^{1-5}$ related to the different processes (electron impact emission and attachment) involved in the production of the positive and negative ions, reveal a small disparity in the two temperatures, and this should in principle allow for an electrostatic acoustic mode to develop.

The very recent successful production of {\em pair-hydrogen plasma}$^5$ is even more attractive because of the small ion mass and the obvious consequences related to this. This indicates that the experimental and analytical efforts in this particular field will increase even further. Therefore, in what follows we discuss various aspects of the stability of waves in pair-ion plasmas, using  some limits  and parameters that are not necessarily always directed only to the presently available experimental results. This study includes various limits of the  plasma density, and the presence of electrons which may naturally appear in such a plasma or can be added for the sake of scientific curiosity to check the effects introduced by their presence.

\vspace{0.7cm}

\noindent{\bf II. \,\,\, MODEL}

\vspace{0.7cm}

 In the experiments$^{1-5}$ due to the specific filtering mentioned above, only pair ions are collected in the chamber, and it is believed that electrons are generally absent.  The arguments supporting this opinion are convincing and, hence, we accept the mentioned filtering as a fact. Yet, electrons may be added to the pair-ion plasma in order to check the effect of their presence on the stability of the plasma. On the other hand, we note that electrons, being attached to the negative ions, still enter the chamber indirectly. Moreover, collisions enable charge exchange between the ions, formally described by a process of the type   $A^+ + B^-\rightarrow A + B$,  and resulting in a number of neutrals in the system. The  electrons can also  be released from the negative ions in any of the collisions of the type $B^- + B\rightarrow B + B + e^-$,  $B^- + A \rightarrow  B + A  + e^-$,  and $A^+ + B^-\rightarrow A^+ + B + e^-$. As a result,  a mixture of pair-ions (because in the present case $A\equiv B\equiv C_{60}$), electrons, and neutrals may be expected. Hence, the momentum equations for ions and electrons, that will be used below, will include the friction force  between different plasma species. According to the given scenario, the presence of electrons should be related to the presence of neutrals. Yet, the neutrals are not expected to play a significant role in the system because in terms of electron collisions, roughly speaking, an ion is equivalent to $3.4 \cdot 10^5 (300/T_e)^2$ neutral atoms.$^{16}$ So collisions between charged species are usually dominant even when the number density of neutrals $n_n$ by far exceeds the number density of ions.   The situation is quite opposite with the presence of electrons. As a matter of fact, even a small amount of electrons can drastically change the dispersive properties of the modes in multi-ion plasmas, and indeed introduce new ones.

 We note also that the spatial variation of the density in both the radial and the axial directions has been experimentally detected,$^{1}$ with the radial density profile being Gaussian. In view of the very different axial and radial lengths of the chamber, the axial density inhomogeneity may be neglected without losing any essential effect. Such equilibrium density gradients are a rather common feature of various laboratory plasmas.$^{17-19}$

 In the experiments 2-4 the modes are initiated by a cylindric exciter at one end of the chamber, and   they also  appear  to propagate {\em  almost parallel} to the magnetic field vector,  i.e., they {\em may} include a  (small) wave number component in the perpendicular direction as well, as commented in Ref.~4.

\vspace{0.4cm}

\noindent{\bf A. Unmagnetized ions}

\vspace{0.4cm}

 We proceed  by introducing  high frequency electrostatic perturbations  propagating obliquely to the equilibrium magnetic field vector $B_0=B_0\vec e_z$.  In the frequency limit $\Omega_e >\omega \gg \Omega_i$, the ions are un-magnetized, the Lorentz force in the ion momentum equation can be omitted, and their perpendicular  (with respect to the magnetic field vector) response to the perturbed electric field is similar to the case of an ordinary ion sound  wave. In principle, the ions will remain un-magnetized even for frequencies below the ion gyro-frequency (that are in fact also experimentally measured)  provided that the perpendicular wave-length is below the ion gyro-radius. The momentum equation for ions of type $j$ is as follows
\be
m_j n_j\frac{\partial \vec v_j}{\partial t}= - n_j q_j \nabla \phi - \kappa T_j \nabla n_j - m_j n_j \nu_j \vec v_j. \label{e1}
\ee
The electron dynamics is described by the  momentum equation
\be
m_en_e \left[\frac{\partial \vec v_e}{\partial t} + (\vec v_e\cdot\nabla)\vec
v_e\right] = e n_e\nabla \phi  - e n_e \vec v_e\times \vec B
 - \kappa T_e \nabla n_e
- m_e n_e \nu_e\vec v_e, \label{e3}
\ee
and we also use the appropriate continuity equation for all species
\be
\frac{\partial n_j}{\partial t}+ \nabla\cdot(n_j \vec v_j)=0. \label{e2}
\ee
Here, the electron collision frequency is in principle the  sum $\nu_e=\nu_{ei}+ \nu_{en}$.

From Eqs.~(\ref{e1}) and (\ref{e2}) we get,$^{20}$ for ion perturbations of the form $\sim f(x) \exp(-i \omega t + i \vec k\cdot \vec r)$, where $\vec r$ denotes an arbitrary direction with respect to $\vec B_0$,
\be
n_{j1}=\frac{n_{j0} q_j k^2 }{m_j} \frac{\phi_1}{\omega^2 + i \omega \nu_i - k^2 \vtj^2}, \label{e4}
\ee
where $\vtj^2=\kappa T_j/m_j$.
The electron dynamics in the perpendicular direction  is described by
the following recurrent formula for the velocity
\[
v_{e\bot}= \frac{1}{B_z}\vec e_z\times \nabla_\bot  \phi +\frac{\nu_e}{\Omega_e}
\frac{\nabla_\bot \phi}{B_z} - \frac{v_{\sss{T} e}^2\nu_e}{\Omega_e^2} \frac{\nabla_\bot
n_e}{n_e}  -  \frac{v_{\sss{T} e}^2}{\Omega_e} \vec e_z\times \frac{\nabla_\bot
n_e}{n_e}
\]
\be
   -
 \frac{1}{\Omega_e} \left(\ddt + \vec v_e\cdot\nabla\right)\vec
e_z\times \vec v_{e\bot} - \frac{\nu_e}{\Omega_e^2}\left(\ddt + \vec
v_e\cdot\nabla\right) \vec v_{e\bot}.  \label{e3a} \ee
The  parallel perturbed electron velocity is given by
 \be
v_{ez1}= -\frac{e k_z}{m_e \omega_1} \phi_1 + \frac{v_{{\sss T} e}^2
k_z}{\omega_1} \frac{n_{e1}}{n_{e0}}. \label{e2a} \ee
Here, $\omega_1 = \omega_0 + i \nu_e$,  $\omega_0=\omega- k_y v_{e0}$,  and
$\vec v_{e0}\approx - (\vec e_z\times \nabla_\bot  n_{e0}/n_{e0})\vte^2/\Omega_e$ is the electron diamagnetic drift.  We have used $\nu_e\ll \Omega_e$ in the electron diamagnetic drift and in the  last term in Eq.~(\ref{e3a}), and we work in the usual drift limit for electrons, with a small equilibrium density gradient, and in a local approximation implying $|d/dx| \ll |k_y|$.

\vspace{0.7cm}

\noindent{\bf III. \,\,\, MODES IN COLD PLASMA}

\vspace{0.7cm}

In order to check for the most basic modes, here we omit collisions and thermal effects, and discuss first the case of a two-component {\em electron-ion plasma}. The simplified  ion equation (i.e.\ without thermal terms, \ref{e4}) is used,
while  the electron density perturbation becomes
$n_{e1}= [n_{e0} k_y^2/(\Omega_e B_0) - e n_{e0} k_z^2/(m_e\omega^2)- k_y n_{e0}'/(\omega B_0)] \phi_1$.
From  the quasi-neutrality condition we obtain  a modified lower hybrid mode, viz.
\be
\omega^2- \frac{\Omega_e n_{e0}'}{k_y n_{e0}} \omega - \frac{k_z^2 \Omega_e^2}{k_y^2} \left(1+ \frac{\Omega_i k^2}{\Omega_e k_z^2}\right)=0.
\label{lh}
\ee
Using the Poisson equation instead of the quasi-neutrality, we obtain
\be
\omega^2\left(1+ \frac{\omega_{pe}^2 k_y^2}{\Omega_e^2 k^2}\right)  - \frac{k_y \omega_{pe}^2 n_{e0}'}{ k^2 \Omega_e n_{e0}} \, \omega -  \omega_{pi}^2 - \frac{\omega_{pe}^2 k_z^2}{k^2} =0. \label{ep}
\ee
Comparing this to Eq.~(\ref{lh}), it is seen that the quasi-neutrality in the perturbed state is equivalent to assuming $\omega_{pe}^2k_y^2/(\Omega_e^2k^2)\gg 1$.
For perpendicular propagation without a density gradient, Eqs.~(\ref{lh}) and (\ref{ep}) yield the lower-hybrid and the ion plasma waves $\omega^2=\Omega_e \Omega_i$ and $\omega^2=\omega_{pi}^2/(1+ \omega_{pe}^2/\Omega_e^2)$, respectively.

In view of the difference in mass, the last three terms in Eq.~(\ref{ep}) can not always be of the same order. An interesting case for which the mode behavior is determined by the  density gradient  term, yielding a backward solution $\partial \omega/\partial k_z<0$, is obtained by the following set of parameters. We take  $B_0=0.1\;$T, $L_n= n_{e0}/n_{e0}'$, where $n_{e0}(r)=N_0 \exp(-r^2/a^2)$, so that locally  $|L_n|= a^2/(2r)$.  If $a=r_0$, or $a=2 r_0$, where  $r_0=4\;$cm is the plasma radius, at $r=r_0$ this gives for the density $0.37$ and $0.78$ times its value at the axis, respectively.
We choose  $L_n=8\;$mm,  and take  three values for the plasma number density, omitting the index $e$ for a quasi-neutral equilibrium density: $n_0=10^{16}\;$m$^{-3}$, $n_0=5\cdot 10^{15}\;$m$^{-3}$, and $n_0=10^{15}\;$m$^{-3}$. The results are  presented in Fig~1. Here, the wave frequency has a value much above the ion gyro frequency, but below the electron gyro frequency,  and the parallel wave length $\lambda_z$ is in the range $10 - 100\;$cm. In the poloidal $y$ (or $\theta$) direction we have  fixed one wave length only, i.e.,  $m=1$. The decreasing (full)  line with $k_z$ (backward mode)  represents the solution of Eq.~(\ref{ep}) for the case of a dominant  density gradient term.  For a lower plasma density (dotted line) we have a direct  mode $\partial \omega/\partial k_z>0$. In a cylindric geometry, the given solutions describe twisted waves, i.e., those  traveling both along the magnetic field lines and in the $\theta$-direction. The backward mode  has a behavior similar to the observed IF wave, yet in the much higher frequency domain.

The same effect, i.e., the  transition from a backward to a direct mode, can be achieved and controlled by changing the poloidal wave length $\lambda_y$. This can easily be checked by setting $n_0=10^{16}\;$m$^{-3}$, and taking $m=10$, which yields a direct mode.

It is worth mentioning that this transition from forward to backward modes by increasing the pressure, has in fact  been  observed experimentally in rare gases as was reported long ago in Ref.~21, but without any explanation about the source of that behavior.

We stress also that {\em in the presence of an additional,  negatively charged ion species of the same mass, the mode behavior does not change}. This is because the case  presented here is an interplay between the two  dominant terms, i.e., the  density gradient term and the last term in Eq.~(\ref{ep}).

\vspace{0.7cm}

\noindent{\bf IV. \,\,\, GROWING MODES IN COLLISIONAL INHOMOGENEOUS  PLASMA}

\vspace{0.7cm}

Using Eqs.~(\ref{e3a}) \& (\ref{e2a}), from the electron continuity equation we obtain:
\be
\frac{n_{e1}}{n_{e0}}= \frac{\omega_* - \vte^2 k_z^2 [1-\alpha \omega_1 \omega_\alpha k_y^2/(k_z^2 \Omega_e^2)]/\omega_1}{
\omega-  \vte^2 k_z^2 [1-\alpha  \omega_1 \omega_\alpha k_y^2/(k_z^2 \Omega_e^2)]/\omega_1} \,\,\frac{ e \phi_1}{\kappa T_e}.
\label{e2b}
\ee
Here, $\omega_\alpha = \omega_0 \alpha + i \nu_e$,  $\alpha=1/(1+\nu_e^2/\Omega_e^2)$, $\omega_*=-k_y \kappa T_e n_{e0}'/(e B_0 n_{e0})$. Note that  $\omega_0$  appears in both  $\omega_\alpha$ and  $\omega_1$ from the left-hand side of the electron momentum equation, i.e.,   as the finite electron mass effect.

\vspace{0.4cm}

\noindent{\bf A. Massless hot electrons}

\vspace{0.4cm}

In the limit $|\omega_0|< \nu_e$,  from Eq.~(\ref{e2b}) we have$^{22}$
\be
\frac{n_{e1}}{n_{e0}}=\frac{\omega_* + i D_e}{\omega+ i De}\, \frac{ e \phi_1}{\kappa T_e}, \quad
D_e=\frac{k_z^2 \vte^2}{\nu_e} + \rho_e^2 k_y^2 \nu_e, \quad \rho_e=\frac{\vte}{\Omega_e}.
\label{e5}
\ee
The second term in $D_e$ is  small compared to the first one provided that  $k_y^2 \nu_e^2/(k_z^2 \Omega_e^2)\ll 1$, implying a dominant effect of collisions on the parallel electron dynamics. In this case,  Eq.~(\ref{e5}) coincides with the corresponding expression from  Ref.~23. The same limit holds when the electron mass corrections  are retained.$^{20,24}$

Using Eq.~(\ref{e5}) instead of Eq.~(\ref{e2b}), with  the   quasi-neutrality in the equilibrium and the Poisson equation in the perturbed state, in the case of the pair-ion electron  plasma with $q_a=- q_b=e$,  we obtain  the dispersion equation
\be
1=\frac{\omega_{pa}^2}{\omega^2 +
 i \omega \nu_a - k^2 \vta^2} + \frac{\omega_{pb}^2}{\omega^2 + i \omega \nu_b - k^2 \vtb^2} -\frac{\omega_{pe}^2}{k^2 \vte^2}\frac{\omega_* + i D_e}{\omega+ i D_e}. \label{e6}
\ee
Here, $\omega_{pj}^2= e^2 n_{j0}/(\varepsilon_0 m_j)$. Below we discuss several particular cases.

\begin{description}
\item{a)} Without electrons and for a collision-less, pure pair-ion plasma with  $m_a=m_b$, and  for modes propagating along the magnetic field vector,   Eq.~(\ref{e6}) yields
\be
\omega^2= \omega_p^2+ \frac{k_z^2 (\vta^2 + \vtb^2)}{2} \pm \left[\omega_p^4+ \frac{k_z^4(\vta^2-\vtb^2)^2}{4}\right]^{1/2}. \label{two}
\ee
In the limit of a small $k_z$ (and/or a small difference between  the two temperatures) this describes the ion plasma wave $\omega^2\simeq 2 \omega_p^2 + k_z^2 (\vta^2+ \vtb^2)/2$, and the ion sound mode $\omega^2\simeq   k_z^2 (\vta^2+ \vtb^2)/2$ when $T_a\neq T_b$.  It is tempting to set $T_a=T_b$ here, however, strictly speaking, the sound mode loses its electrostatic nature in this case and it becomes an ordinary gas acoustic mode (or thermal mode) involving only the pressure perturbations.

Specific conditions in the experiments,$^{1-5}$ related to different processes  involved in  the production of positive and negative ions (electron impact emission and attachment), reveal a small disparity in the two temperatures  $T_a \neq T_b$.  Hence, solving Eq.~(\ref{two}) with the parameters  $n_{i0}= 10^{13}\;$m$^{-3}$,  $T_a=0.5\;$eV, $T_b=0.3\;$eV, and for $k_z$ in the range $6 - 125$ m$^{-1}$,  yields the ion plasma wave with frequency  $\simeq 220\;$kHz, and the ion sound wave in the range $1.5 - 29\;$kHz. The sound mode in the experiments$^{2-5}$ appears divided into two branches. The values for the ion sound which follow from Eq.~(\ref{two}) are in good  agreement with the lower experimental branch. However, for explaining the upper experimental branch the presence of electrons may be needed. In view of  a large difference in mass, the electrons are normally described by the Boltzmann distribution. Yet, adding such a contribution of electrons only yields corrections that are of importance at spatial scales of the order of the electron Debye scale $\lambda_{de}=\vte/\omega_{pe}$. Therefore, extra effects are needed, like  those that follow from Eq.~(\ref{e2b}), and from the last term in (\ref{e6}).

\item{b)} In an electron-ion  ($n_b=0$) quasi-neutral plasma ($k^2 \lambda_{de}^2\ll 1$) with cold ions,   from Eq.~(\ref{e6}) we have
\be
 \omega^2=k^2 c_s^2 \left(1+ \frac{\omega-\omega_*}{\omega_* + i D_e}\right),\label{e7a}
 \ee
 describing a high-frequency  (as assumed $\omega>\Omega_i$) modified ion acoustic mode that may be unstable  due to the density gradient and electron collisions. The numerical solution for the growth rate of the strongly growing mode  (\ref{e7a}) is presented in Fig.~2, for a relatively high (necessary to have all conditions satisfied) plasma number density $n_{e0}= 10^{19}\;$m$^{-3}$, the magnetic field $B_0=0.2\;$T, for cold ions, and $T_e=5\;$eV, $L_n=2\;$cm, and for the three values $\lambda_y \simeq 1, \,2,\, 5\;$mm. For these parameters and the  given
$\lambda_y$, we have $\omega_*=7.8\cdot 10^6, \,4.7\cdot 10^6,\, 1.6\cdot 10^6 \;$Hz,   and the corresponding modified sound wave frequency
$\omega\simeq  5 \cdot 10^6,\, 3 \cdot 10^6, \,1 \cdot 10^6\;$Hz, respectively. The mode is growing for $\omega_*>\omega$. Here $m_i=720 \, m_p$, hence  $\Omega_i\ll \omega$ and the ions are un-magnetized. In the same time, $k^2 \lambda_{de}^2\ll 1$ for all three cases discussed, so that the quasi-neutrality is well satisfied, while  $\omega_0/\nu_e\ll 1$ so that the electron inertia is negligible. The FEM (finite electron mass) line is described  further in the text.

\item{c)} In the case of pair-ion electron plasma with the Boltzmann electron response (that implies $\nu_e\sim \omega$ and $\omega/k\ll \vte$),
   the last term in Eq.~(\ref{e6}) is to be  replaced with $\omega_{pe}^2/(k^2 \vte^2)$.  Setting  $T_a=T_b=T$ and $m_a=m_b=m$ yields only one pair of solutions $\omega^2=k^2 c_s^2 [T/T_e + (n_{a0}+ n_{b0})/n_{e0}]$ for $k^2 \lambda_e^2\ll 1$, and
 $\omega^2=(\omega_{pa}^2 + \omega_{pb}^2)k^2 \lambda_e^2/(1+ k^2 \lambda_e^2)  + k^2 v_{{\sss T}}^2$ for $k^2 \lambda_e^2\geq  1$, where $c_s^2=\kappa T_e/m$, $v_{{\sss T}}^2= \kappa T/m$.

Otherwise, from Eq.~(\ref{e6}) for cold ions  we have  the third order  dispersion equation
 \be
 \omega^3 + \left(\frac{\omega_*}{k^2\lambda_{de}^2}+ i s D_e\right) \omega^2 - \omega_{pa}^2 \omega - i D_e \omega_{pa}^2=0. \label{e7}
\ee
Here, $s=1+1/(k^2\lambda_{de}^2)$.
Unstable solutions may be detected by using the generalized Hurwitz method for polynomials with
complex coefficients.$^{25,26}$. According to this, for a polynomial of the degree $m$, of the form  $(a_0+ i b_0)x^m + (a_1 + i b_1) x^{m-1} + \ldots + (a_m + i b_m)=0$, one makes the sequence of $m+1$ numbers $c_0=a_0$, $c_1=a_1$,
$\ldots$, $c_r$, $\ldots$, where $r$ goes to $m$, and where
\[
c_r=(-1)^{r(r-1)/2}
 \left|
\begin{array}{cccccc}
  a_1 & a_0 & 0 & 0 & 0 & \cdot \\
  -b_2 & -b1 & a_1 & a_0 &0 & \cdot \\
  a_3 & a_2 & b_2 & b_1 & a_1 & \cdot \\
  -b_4 & -b_3 & a_3 & a_2 & -b_2 & \cdot \\
 a_5 & a_4 & b_4 & b_3 & a_3 & \cdot \\
  \cdot & \cdot & \cdot & \cdot & \cdot & \cdot \\
  \cdot & \cdot & \cdot& \cdot & \cdot& \cdot \\
  a_{2r -1} & a_{2r -2} & b_{2r-2} & b_{2r-3} & a_{2r - 3} & \cdot
\end{array}
\right|. \vspace{0.3cm}
\]

The number of roots with positive real parts
equals the number of sign changes in the sequence $c_j$. A sufficient
instability condition is that any of the $c_r$ has a negative sign.  Applying this to Eq.~(\ref{e7}), we find: $c_0=1$, $c_1=\omega_*/(k^2 \lambda_{de}^2)$, $c_2=-\omega_{pa}^2 \omega_*^2/(k^2 \lambda_{de}^2)^2$, $c_3=- D_e^2 \omega_*^3 \omega_{pa}^4/(k^6 \lambda_{de}^6)$.
This indicates that, for $\omega_*>0$,  Eq.~(\ref{e7}) has one mode propagating in the positive direction, and two solutions propagating in the opposite direction, and there exists at least one growing solution. Another possibility is  $\omega_*<0$. According to experimental conditions $n_{e0}'<0$. Therefore in this case we have $k_y<0$, there are two sign changes in the series $c_j$, hence two positive solutions for $\omega$, and we still have at least one growing solution because $c_1, c_2<0$.

\item{d)} The  case a)   does not change considerably in the presence of  boltzmannian electrons in the collision-less plasma and for the same parallel propagation when we have
\[
\omega^2=\frac{\omega_{pa}^2 + \omega_{pb}^2}{2 s} + \frac{k_z^2}{2} (\vta^2 + \vtb^2)
\]
\[
 \pm \frac{1}{2}
\left\{\frac{(\omega_{pa}^2 + \omega_{pb}^2)^2}{s^2} +   k_z^2 (\vta^2 - \vtb^2)\left[ \frac{2(\omega_{pa}^2 - \omega_{pb}^2)}{s} +
k_z^2 (\vta^2 - \vtb^2)\right]\right\}^{1/2}.
\]
This is just a modified form of Eq.~(\ref{two}).
\end{description}

\vspace{0.4cm}

\noindent{\bf B. Full dispersion equation}

\vspace{0.4cm}

Keeping the full form of Eq.~(\ref{e2b}), with the help of the Poisson equation, the dispersion equation becomes
\[
1= \frac{\omega_{pa}^2}{\omega^2 +
 i \omega \nu_a - k^2 \vta^2} + \frac{\omega_{pb}^2}{\omega^2 + i \omega \nu_b - k^2 \vtb^2}
 \]
 \be
 - \frac{1}{k^2 \lambda_e^2} \frac{\omega_* - \vte^2 k_z^2 [1-\alpha \omega_1 \omega_\alpha k_y^2/(k_z^2 \Omega_e^2)]/\omega_1}{
\omega-  \vte^2 k_z^2 [1-\alpha  \omega_1 \omega_\alpha k_y^2/(k_z^2 \Omega_e^2)]/\omega_1}. \label{full}
\ee
Eq.~(\ref{full}) will be solved numerically for some sets of parameter values.

We first check the effects of the finite electron mass (FEM), and solve Eq.~(\ref{full}) for the electron-ion case ($n_{b0}=0$) with the same parameters as in Fig.~2, for cold ions,  and for the case $k_y=3770\;$m$^{-1}$ (dashed line). The sound wave frequency appears to be slightly larger, i.e., in the range $3.1 - 3.2\;$MHz, while the growth rate is reduced, given by the lower dashed line in Fig.~2.

For the pair-ion electron plasma, Eq.~(\ref{full}) is solved for collision-less ions in terms of $n_{b0}/n_{a0}$, and the result for the acoustic mode  is presented in Fig.~3 for the same parameters as in Fig.~2 and for fixed values of $k_z=18.8\;$m$^{-1}$ and $k_y=3770\;$m$^{-1}$. The driving source for the instability is the density gradient in combination with electron collisions. Hence, for a reduced electron number density the instability vanishes; in the present case this happens at  $n_{b0}/n_{a0}\simeq 0.4$.
The other pair of solutions,  that is in the range of the ion plasma frequency and the lower hybrid frequency, appears strongly damped, with $\omega_r$ being of the same order as $|\omega_i|$ and it is consequently of no interest.

For much lower density, e.g., setting    $n_{a0}= 10^{14}\;$m$^{-3}$ and varying $n_{b0}/n_{a0}$, we obtain  two  direct modes with frequencies close to $10^6\;$Hz and $2\cdot 10^7\;$Hz, propagating almost without change, the former very weakly growing and the latter being almost undamped.

In the case of different temperatures of two pair-ion species (like in the reported cases in Refs. 2-5), there appears an additional acoustic mode. This is checked by setting $T_a=0.5\;$eV, $T_b=0.3\;$eV for   $n_{a0}= 10^{19}\;$m$^{-3}$.  The other parameters are the same as in Fig.~3. In total, there are three pairs of solutions: one high frequency $\sim 10^7\;$Hz but highly damped mode mentioned above, the slightly modified acoustic mode from Fig.~3, and the third, weakly growing acoustic mode presented in Fig.~4.

We stress that the negative solutions (that are not presented in the figures above) are in fact  not counterpart solutions in the strict sense because of the asymmetry caused by the density gradient.

\vspace{0.7cm}

\noindent{\bf V. \,\,\, WEAKLY MAGNETIZED IONS}

\vspace{0.7cm}

In the experiments$^{2-5}$ the ions in the equilibrium are magnetized. However, for perturbations with frequencies above the ion gyro-frequency, the ions are un-magnetized, yet,  for not so high frequency oscillations they can feel the effects of the magnetic field. Hence, in Eq.~(\ref{e1}) we add the Lorentz force. The equilibrium diamagnetic drift velocity for the two ion species is given by $\vec v_{j0}=(\vec e_z \times \nabla n_{j0}/n_{j0})\kappa T_j/(q_j B_0)$. For the given experimental parameters ($T_a=0.5\;$eV, $T_b=0.3\;$eV) and assuming $L_n$ of the order of 1~cm, the two velocities are of the order of $10^2\;$m/s. For perpendicular wavelengths of the order of 1~cm,  the Doppler shift in the frequency, $k_y v_{j0}$,  becomes  of the order  $10^5\;$Hz, and can therefore introduce visible effects in the mode behavior. Assuming a negative density gradient, we have  a negative $v_{a0}\vec e_x$ and $\vec v_{b0}=- \vec v_{a0} T_b/T_a$.
The velocity components for the ion  species $a$ become:
\[
v_{ax1}=\frac{i}{\omega_{a0}^2-\Omega^2}\left[k_y \Omega^2 \frac{\phi_1}{B_0} + \vta^2 \left( k_y \Omega - \omega_{a0} \left|\frac{n_{a0}'}{n_{a0}}\right| \right)\frac{n_{a1}}{n_{a0}} \right],
\]
\[
v_{ay1}= \frac{1}{\omega_{a0}^2-\Omega^2} \left[k_y \omega_{a0} \Omega \frac{\phi_1}{B_0} + \vta^2 \left( k_y \Omega - \omega_{a0} \left|\frac{n_{a0}'}{n_{a0}}\right| \right)\frac{n_{a1}}{n_{a0}} \right],
\]
\[
v_{az1}=\frac{k_z \Omega}{\omega_{a0}} \frac{\phi_1}{B_0} + \frac{k_z \vta^2}{\omega_{a0}} \frac{n_{a1}}{n_{a0}}.
\]
Here, $\omega_{a0}=\omega- k_y v_{a0}$.  In the corresponding equations for the ion species $b$,  the term  $\Omega$ is to be taken with the minus sign.

The continuity equation  $n_{j1}/n_{j0}=(k_y v_{jy1} + k_z v_{jz1} + i v_{jx1}|n_0'/n_0| )/\omega_{j0}$ for the two species, and the Poisson equation, yield  the dispersion equation for the modes in an inhomogeneous pair-ion plasma:
\[
\frac{k^2}{\omega_p^2}=
\frac{\frac{\displaystyle{k_y^2}}{\displaystyle{\omega_{a0}^2-\Omega^2}} + \frac{\displaystyle{k_z^2}}{\displaystyle{\omega_{a0}^2}} - \left|\frac{\displaystyle{n_{a0}'}}{\displaystyle{n_{a0}}}\right|  \frac{\displaystyle{k_y \Omega}}{\displaystyle{\omega_{a0} (\omega_{a0}^2- \Omega^2)}}}
{
1-\frac{\displaystyle{k_z^2 \vta^2}}{\displaystyle{\omega_{a0}^2}} - \frac{\displaystyle{k_y^2 \vta^2}}{\displaystyle{\omega_{a0}^2-\Omega^2}}\left(1- \frac{\displaystyle{\Omega}}{\displaystyle{\omega_{a0}}} \frac{\displaystyle{1}}{\displaystyle{k_y}} \left|\frac{\displaystyle{n_{a0}'}}{\displaystyle{n_{a0}}}\right|\right) +  \left|\frac{\displaystyle{n_{a0}'}}{\displaystyle{n_{a0}}}\right| \frac{\displaystyle{\vta^2}}{\displaystyle{\omega_{a0}^2-\Omega^2}}\left(\frac{\displaystyle{k_y \Omega}}{\displaystyle{\omega_{a0}}} - \left|\frac{\displaystyle{n_{a0}'}}{\displaystyle{n_{a0}}}\right|\right)}
\]
\be
+ \frac{\frac{\displaystyle{k_y^2}}{\displaystyle{\omega_{b0}^2-\Omega^2}} + \frac{\displaystyle{k_z^2}}{\displaystyle{\omega_{b0}^2}} + \left|\frac{\displaystyle{n_{a0}'}}{\displaystyle{n_{a0}}}\right|  \frac{\displaystyle{k_y \Omega}}{\displaystyle{\omega_{b0} (\omega_{b0}^2- \Omega^2)}}}
{
1-\frac{\displaystyle{k_z^2 \vtb^2}}{\displaystyle{\omega_{b0}^2}} - \frac{\displaystyle{k_y^2 \vtb^2}}{\displaystyle{\omega_{b0}^2-\Omega^2}}\left(1+ \frac{\displaystyle{\Omega}}{\displaystyle{\omega_{b0}}} \frac{\displaystyle{1}}{\displaystyle{k_y}} \left|\frac{\displaystyle{n_{a0}'}}{\displaystyle{n_{a0}}}\right|\right) -  \left|\frac{\displaystyle{n_{a0}'}}{\displaystyle{n_{a0}}}\right| \frac{\displaystyle{\vtb^2}}{\displaystyle{\omega_{b0}^2-\Omega^2}}\left(\frac{\displaystyle{k_y \Omega}}{\displaystyle{\omega_{b0}}} + \left|\frac{\displaystyle{n_{a0}'}}{\displaystyle{n_{a0}}}\right|\right)}.\label{pair}
\ee
In the absence of electrons, the density gradients and number densities are the same, and $\omega_p^2=e^2n_0/(\varepsilon_0 m)$.
Various limits of Eq.~(\ref{pair}) can easily be discussed, yielding the previously discussed modes. For example,  in the absence of a density gradient and for strictly perpendicularly propagating  oscillations in a quasi-neutral plasma, from Eq.~(\ref{pair}) we have just the ion cyclotron mode $\omega^2=\Omega^2 + k^2 (\vta^2 + \vtb^2)/2$. For strictly parallel propagation in a quasi-neutral plasma we have the acoustic mode $\omega^2=k^2(\vta^2+ \vtb^2)$, otherwise it yields Eq.~(\ref{two}).

A particularly interesting limit is seen also in the case of a cold plasma and for long wavelengths, when the remaining density gradient terms in the numerators in   Eq.~(\ref{pair}) exactly cancel each other, and it  yields electrostatic convective cells $\omega^2=\Omega^2 k_z^2/(k_y^2 + k_z^2)$. This result is completely different compared to the standard  electron-ion plasma which in such a geometry has the drift wave driven by the density gradient.

Eq.~(\ref{pair}) is solved numerically for parameters close to the experimental values. We take $B_0=0.2\;$T, $T_a=0.5\;$eV, $T_b=0.3\;$eV, but with a lower number density $n_{a0}=n_{b0}=10^{12}\;$m$^{-3}$, and we take $L_n=0.8\;$cm, $k_y/(2 \pi)=0.05\;$cm$^{-1}$. The result for modes with positive frequency is presented in Fig.~5. For values of  $k_z$ in the range $0.01 - 1.15$ (in the given units), the ion plasma mode is in fact a backward one (frequency decreases from $14.05$ at $k_z=0.01$ to $12.7$ at $k_z=0.15$), and these values are only slightly below the experimental values that correspond to the  IF mode. However, we can obtain exactly the same frequency as for the observed IF mode by  simply slightly increasing the density to the value $n_{a0}=n_{b0}=3\cdot 10^{12}\;$m$^{-3}$, but then the  decrease with $k_z$ in this case becomes less pronounced, though it still remains there. We suspect that the unexplained experimental IF mode may in fact be  the ion plasma mode presented here, while  the  experimental upper mode (which in Refs.~$^{2-52}$ has been interpreted as the ion plasma mode) may  be an eventual contribution of  electrons, described by  our Eqs.~(\ref{e2b}) and (\ref{e5}). However, these statements  are  speculative as we have not been able to obtain an additional electron mode in the given experimental range, i.e., at frequencies about 30 and higher (in the units given in Fig.~5).

Due to the density gradient there is a coupling between modes $b$ and $c$, in the range  $k_z \approx 0.032 - 0.07$, with the maximum growth rate $\omega_i$ of the order $\omega_i/\omega_r\simeq 0.2$. The corresponding growth rate, multiplied by 10, is  presented by dashed line in Fig.~5.  The density gradient modifies also the acoustic mode $d$ so that for small values $k_z$ its frequency becomes negative. This all can  easily be demonstrated by switching of the density gradient terms.

We note that the line $b$ has a range of $k_z$ for which it is backward $\partial \omega/\partial k_z<0$, and the same holds for the line $c$.
The frequency of both modes, in the range where they have  backward properties, is in the domain of the low frequency  backward mode observed in the experiment (around 4 in the given units). Though, the agreement with experimental solutions is far from perfect, and the splitting of the lines $b$ and $c$ we obtain for $k_z>0.07$ has indeed not been observed.

\vspace{0.7cm}

\noindent{\bf VI. \,\,\, CONCLUSIONS}

\vspace{0.7cm}

The pair-ion plasma containing heavy fullerene ions (with equal positive and negative charge), that has been created in the laboratory, has attracted a lot of interest in the recent past. The present study deals with some additional properties of such plasmas, e.g., those that result from the eventual  presence of electrons, from plasma inhomogeneity combined  with electron collisions, and from much higher densities.

The arguments given in  Sec.~II show that the presence of electrons can not be excluded, or  that they can be added to the system in order to check the consequences of their presence. In a recent study$^{9}$ some properties of waves in pair ion plasma have been discussed, showing that in fact  they can be used as an indicator of the presence of electrons. Some indications in this direction can also be found in the most recent experimental
results,$^{4,5}$ where the measured sound branch in fact splits into two sub-branches. In the present work, the effects of the density gradient and electron collisions in a relatively high density pair-ion plasma have been explored showing the presence of a backward-propagating high-frequency mode with features similar to the observed IF mode that remained unexplained so far. However, our first backward  mode, determined by the density gradient (see Fig.~1), appears in a  much higher frequency range and it is therefore  unable to explain the observed IF mode. The case of a pure pair-ion plasma, discussed in Sec.~V, reveals that the ion plasma wave behaves as a backward mode in a large range of the parallel wave number, and perhaps may be interpreted as the IF mode observed in the experiment. Yet, the agreement with experimental modes is not good enough to draw a definite conclusion.  Nevertheless, in our view, the plasma features discussed here  represent a step forward in the theory of such a newly created plasma. In the same time we believe that the presented results should  make a good basis for future investigations directed towards the exact explanation of the IF mode.

\vspace{1cm}

\paragraph{Acknowledgements:}
 The  results presented here  are  obtained in the framework of the
projects G.0304.07 (FWO-Vlaanderen), C~90205 (Prodex),  GOA/2004/01
(K.U.Leuven),  and the Interuniversity Attraction Poles Programme -
 Belgian State - Belgian Science Policy.

 \pagebreak

\vfill

\pagebreak

\noindent {\bf Figure captions:}

\begin{description}
\item{1.} The positive solution  of Eq.~(\ref{ep}) for three number densities in electron-ion plasma, showing  the transition from
direct to backward mode. The parameters are given in the text.
\item{2.} The growth rate of the sound mode from Eq.~(\ref{e7a}) for three values of the perpendicular wave lengths. FEM describes the difference due to  electron inertia.
\item{3.} The positive solution of Eq.~(\ref{full}) for a modified ion acoustic mode
as a function of the negative-ion number density, and for equal temperatures of the two ion species.
\item{4.} Additional acoustic mode from Eq.~(\ref{full}) in the case of different temperatures of positive and negative ions, for
$T_a=0.5\;$eV, and  $T_b=0.3\;$eV.
\item{5.} The positive numerical solutions of Eq.~(\ref{pair}) for a pure pair-ion plasma. The growth rate of the coupled modes (multiplied by 10) is given by the dashed line.

\end{description}


\begin{thebibliography}{99}
\bibitem{h1} W. Oohara and R. Hatakeyama, Phys. Rev. Lett. {\bf
91}, 205005 (2003).
\bibitem{h2} W. Oohara and R. Hatakeyama, Phys. Rev. Lett. {\bf
95}, 175003 (2005).
\bibitem{h3} R. Hatakeyama and W. Oohara, Phys. Scripta T {\bf 116}, 101 (2005).
\bibitem{h4} W. Oohara, Y. Kuwabara,  and R. Hatakeyama, Phys. Rev. E {\bf
75}, 056403 (2007).
\bibitem{h5} W. Oohara and R. Hatakeyama, Phys. Plasmas {\bf 14}, 055704 (2007).
\bibitem{v1} J. Vranjes and S. Poedts, Plas. Sources Sci. Tech. {\bf 14}, 485  (2005).
\bibitem{l2} H. Schamel and A. Luque, New J.   Phys.   {\bf 7}, 69 (2005).
\bibitem{s6} A. Luque, H. Schamel, B. Eliasson, and P. K. Shukla, Phys. Plasmas {\bf  12} 122307 (2005).
\bibitem{v2} H. Saleem,  J. Vranjes and S. Poedts,   Phys. Lett. A {\bf 350},
375  (2006).
%
\bibitem{k1}  I. Kourakis, A. Esfandyari-Kalejahi, M. Medhipoor, and P. K. Shukla,  Phys. Plasmas  {\bf 13}, 052117 (2006).
\bibitem{k3}  A. Esfandyari-Kalejahi, I. Kourakis,  and P. K. Shukla,  Phys. Plasmas  {\bf 13}, 122310 (2006).
\bibitem{l1} A. Luque, H. Schamel, B. Elisasson, and P. K. Shukla,  Plas. Phys. Control. Fusion  {\bf 48}, L57 (2006).
\bibitem{s2} H. Saleem, Phys. Plasmas  {\bf 13}, 044502 (2006).
\bibitem{s4} B. Zhao and J. Zheng,  Phys. Plasmas {\bf  14}, 062106 (2007).
\bibitem{s5} I. Kourakis, F. Verheest, and N. F. Cramer, Phys. Plasmas {\bf  14}, 022306 (2007).
\bibitem{rat}  J. A. Ratcliffe, {\em The Magneto-Ionic Theory and its Applications to the Ionosphere}
 (Cambridge University Press, Cambridge, 1959) p. 33.
%
\bibitem{ok1} J. Vranjes, A. Okamoto,  S. Yoshimura,  S. Poedts,
M. Kono, and M. Y. Tanaka,  Phys. Rev. Lett. {\bf 89}, 265002 (2002).
\bibitem{ok2} A. Okamoto,  K. Hara, K. Nagaoka {\em et al.},   Phys. Plasmas {\bf 10}, 2211  (2003).
%
\bibitem{grul} O. Grulke, F. Greiner, T. Klinger, and A. Piel,    Plasma. Phys. Control. Fus. {\bf 43}, 525 (2001).
%
\bibitem{v3} J. Vranjes, M. Y. Tanaka,  and S. Poedts, Phys. Plasmas {\bf 13}, 122103  (2006).
\bibitem{bar} P. J. Barett and P. F. Little, Phys. Rev. Lett. {\bf 14}, 356  (1965).
\bibitem{mih}  A. B. Mikhailovskii, {\em Theory of Plasma Instabilities, vol. 2}
 (Consultants Bureau, New York, 1974) p. 192, and p. 79.
\bibitem{wei} J. Weiland,   {\em Collective Modes in Inhomogeneous
Plasmas} (Institute of Physics Pub., Bristol, 2000) p. 32.
\bibitem{v4} J. Vranjes, B. P. Pandey,  and S. Poedts, Phys. Plasmas {\bf 14}, 032106  (2007).
\bibitem{g}  D. L. Giaretta,   Astron. Astrophys. {\bf  75}, 237 (1979).
\bibitem{v5}  J. Vranjes  and S. Poedts,   Astron. Astrophys. {\bf  458}, 635 (2006).
%

\end{thebibliography}
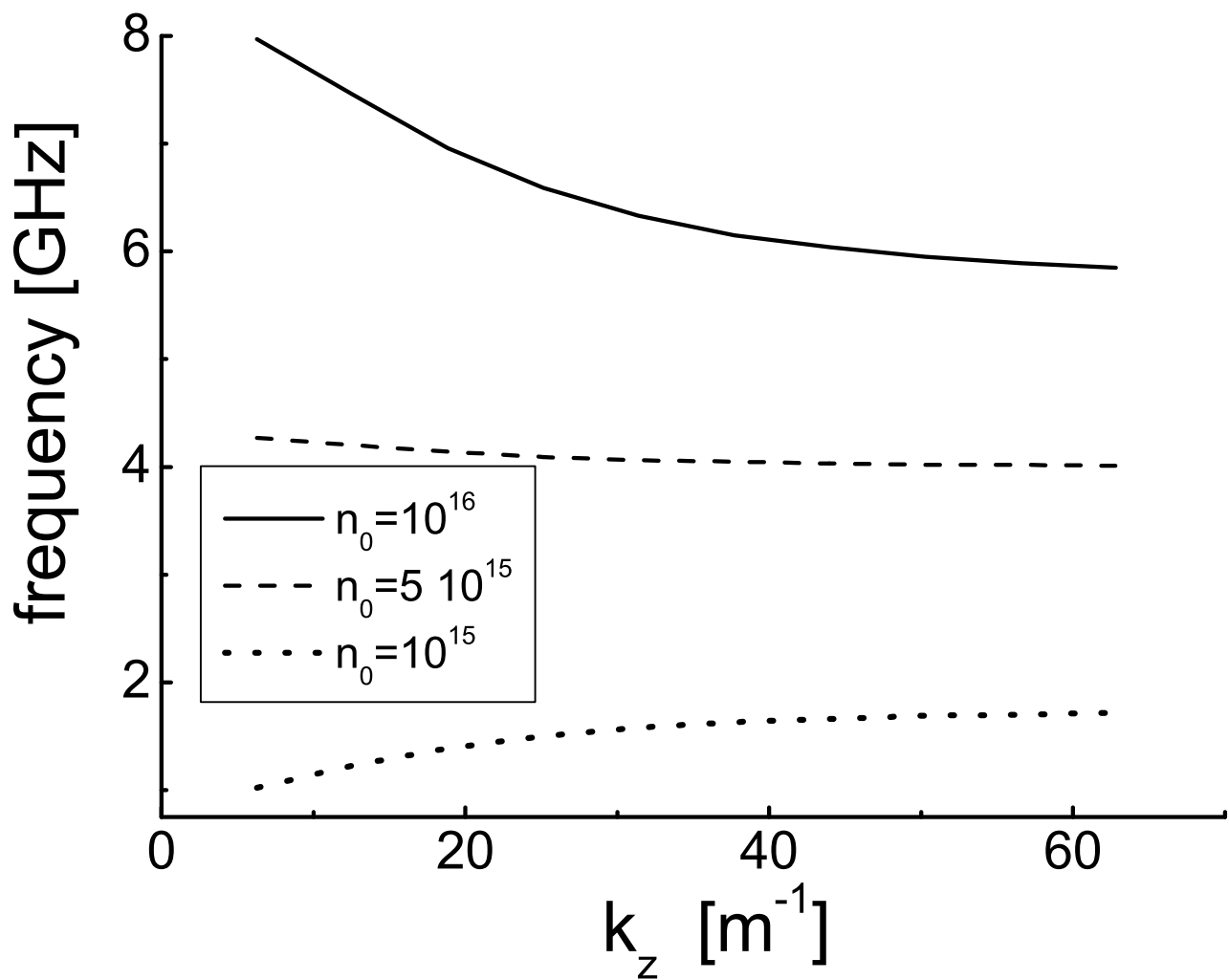
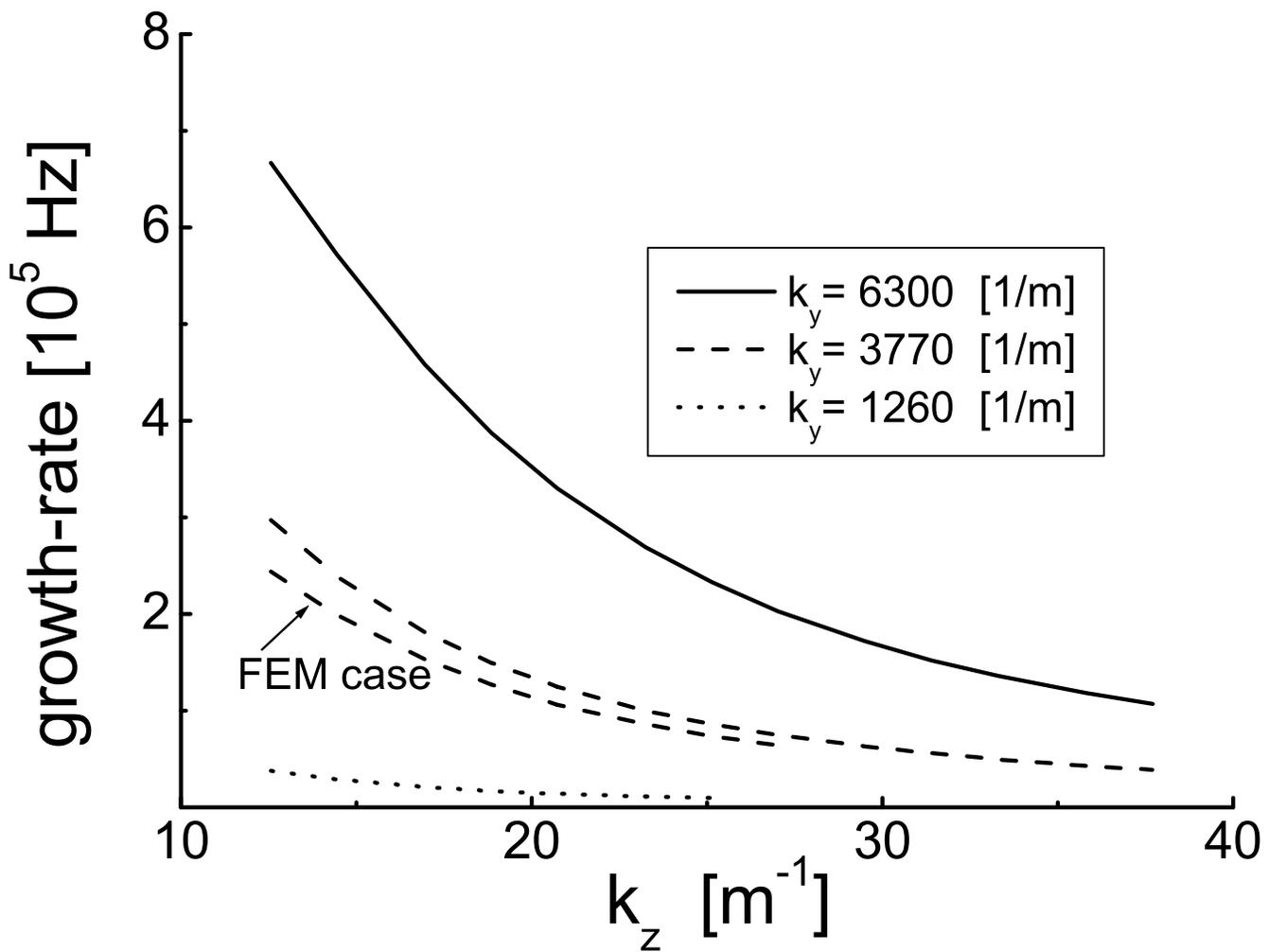
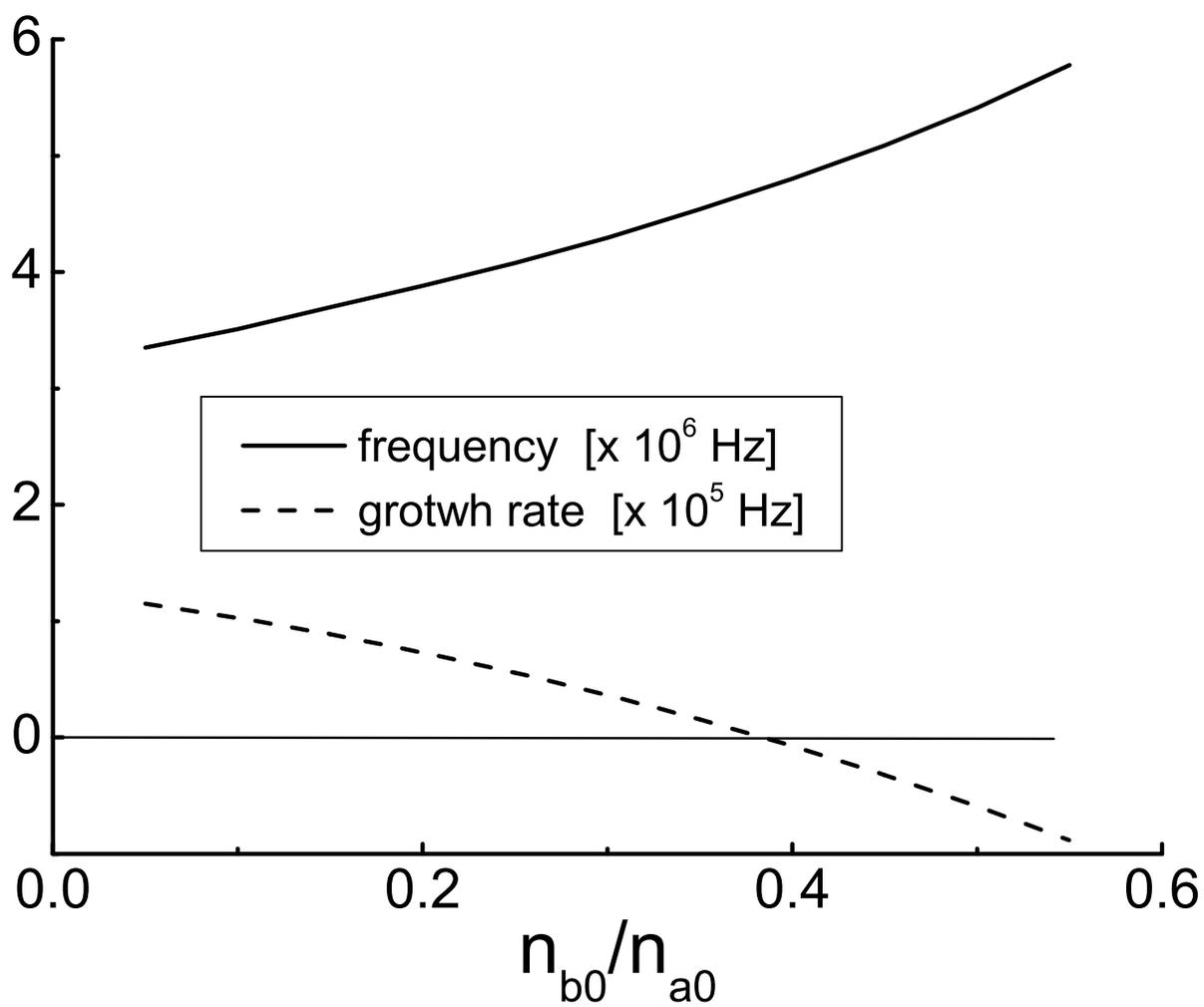
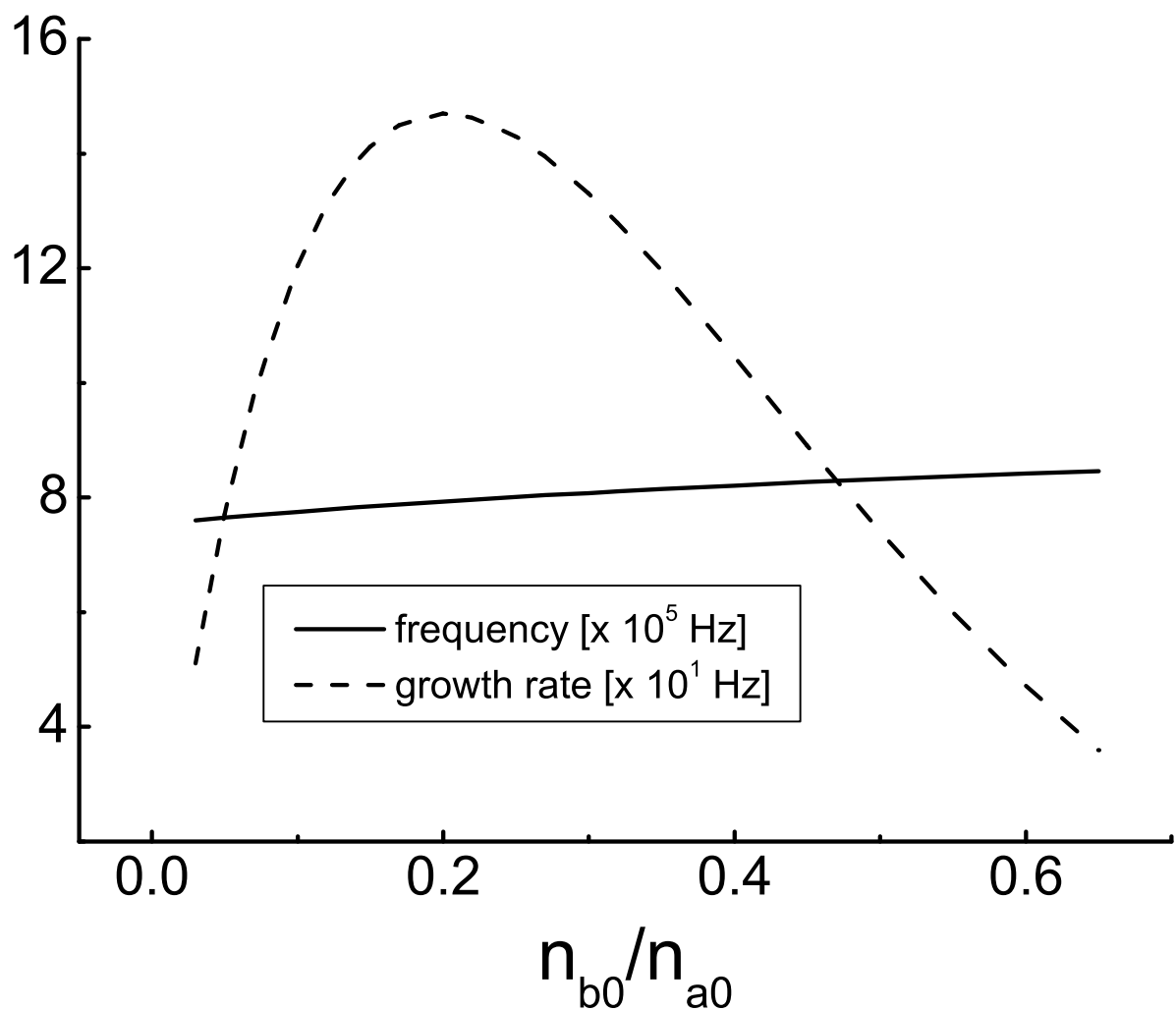
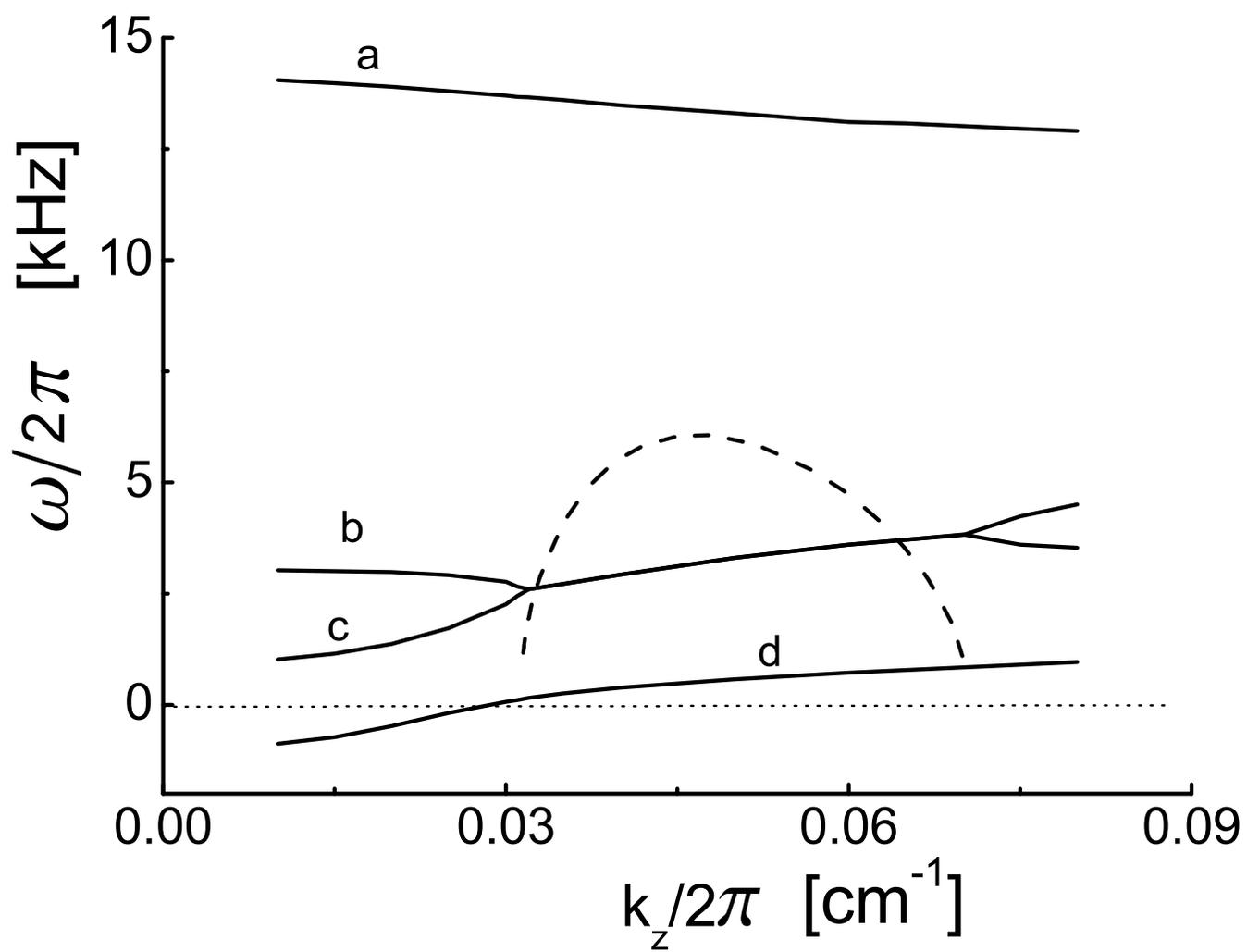
\end{document}